\documentstyle[aps,12pt]{revtex}
\begin {document}
\draft

\title{Geometric phases for neutral and charged particles in a
time-dependent magnetic field\thanks{published in J. Phys. A {\bf 35}
(2002) 377-391.}}
\author{Qiong-gui Lin\thanks{E-mail:
        qg\_lin@163.net, qg\_lin@263.net}}
\address{China Center of Advanced Science and Technology (World
    Laboratory),\\
        P.O.Box 8730, Beijing 100080, People's Republic of China
        \thanks{not for correspondence}\\
        and\\
        Department of Physics, Zhongshan University, Guangzhou
        510275,\\
        People's  Republic of China}

\maketitle
\vfill

\begin{abstract}
\baselineskip 15pt {It is well known that any cyclic solution of a
spin $1/2$ neutral particle moving in an arbitrary magnetic field has
a nonadiabatic geometric phase proportional to the solid angle
subtended by the trace of the spin. For neutral particles with higher
spin, this is true for cyclic solutions with special initial
conditions. For more general cyclic solutions, however, this does not
hold. As an example, we consider the most general solutions of such
particles moving in a rotating magnetic field. If the parameters of
the system are appropriately chosen, all solutions are cyclic. The
nonadiabatic geometric phase and the solid angle are both calculated
explicitly. It turns out that the nonadiabatic geometric phase
contains an extra term in addition to the one proportional to the
solid angle. The extra term vanishes automatically for spin $1/2$.
For higher spin, however, it depends on the initial condition. We
also consider the valence electron of an alkaline atom. For cyclic
solutions with special initial conditions in an arbitrary strong
magnetic field, we prove that the nonadiabatic geometric phase is a
linear combination of the two solid angles subtended by the traces of
the orbit and spin angular momenta. For more general cyclic solutions
in a strong rotating magnetic field, the nonadiabatic geometric phase
also contains extra terms in addition to the linear combination.}
\end{abstract}
\vfill \pacs{03.65.Ta, 03.65.Vf}
\newpage
\baselineskip 15pt

\section{Introduction}               

The motion of spin (especially spin $1/2$) in a rotating magnetic
field is a rather classical problem in quantum mechanics, which was
discussed in the textbook \cite{1}. Nevertheless, the problem has
received much attention in recent years \cite{2,3,4,5,6,7}. The
reason may be that the Schr\"odinger equation for the problem can be
solved analytically, and thus it serves as a good example for
manifesting the notions of adiabatic geometric phase, nonadiabatic
geometric phase for cyclic and noncyclic motions
\cite{8,9,10,11,12,13}. Moreover, it is relevant to some problems in
condensed matter physics \cite{7}.

Cyclic solutions with special initial conditions were widely
discussed in the above cited papers, for both spin $1/2$ and higher
ones. It is well known that the nonadiabatic geometric phase for such
solutions is always proportional to the solid angle subtended by the
trace of the spin (more exactly, the mean value of the spin). Because
the nonadiabatic geometric phase is a geometric object, and because
the result holds for any cyclic solution of spin $1/2$ in an
arbitrarily varying magnetic field \cite{13,14}, one may become
confident that it is also true for higher spin. It is indeed true for
special cyclic solutions in a rotating magnetic field as just
mentioned. For more general cyclic solutions and more general
magnetic fields, however, the result was neither proved nor refuted.
In fact, the nonadiabatic geometric phase for solutions with more
general initial conditions was calculated by some authors only for
spin $1/2$ \cite{7,14}. That for higher spin was, however, not
studied to our knowledge.

In this paper we will consider both neutral and charged particles. In
the next section we consider neutral particles with general spin and
with magnetic moment moving in a rotating magnetic field. The
Schr\"odinger equation for the problem can be solved exactly by
making use of a time-dependent unitary transformation. Solutions with
special initial conditions are cyclic and has been studied in detail
\cite{2}. When the parameters of the system are appropriately chosen,
all solutions are cyclic. These solutions were not discussed in
detail previously. We calculate the nonadiabatic geometric phase for
such solutions. The solid angle subtended by the trace of the spin is
also calculated explicitly. It turns out that the nonadiabatic
geometric phase contains an extra term in addition to the ordinary
one proportional to the solid angle. For spin $1/2$ the extra term
vanishes automatically. This may be the reason why it was not found
previously. For higher spin, however, it depends on the initial
condition and does not vanish in general. At this stage one may
wonder when this extra term does not appear for an arbitrarily
varying magnetic field. This is investigated in Sec. III. We prove
that a sufficient condition is that the initial state is an
eigenstate of ${\bf s}\cdot {\bf e}_0$ where ${\bf s}$ is the spin
operator and ${\bf e}_0$ is some unit vector. Though this conclusion
is known in the literature \cite{layton,gao}, our proof seems more
straightforward and simpler.

In a recent work, we have studied a charged particle moving in a
central potential plus a strong rotating magnetic field \cite{15}. It
can describe the valence electron of an alkaline atom or that of the
hydrogen atom under the influence of the external magnetic field. The
Schr\"odinger equation may be reduced to a Schr\"odinger-like one
with a time-independent effective Hamiltonian by using an explicit
time-dependent unitary transformation. Thus the evolution operator
for the original Schr\"odinger equation was explicitly obtained,
which involves no chronological product. Cyclic solutions are
obtained if one takes the eigenstates of the effective Hamiltonian as
initial states. These eigenstates and the nonadiabatic geometric
phases of the corresponding cyclic solutions were all worked out
explicitly. The nonadiabatic geometric phase turns out to be a linear
combination of the two solid angles subtended by the traces of the
orbit and spin angular momenta. We also studied the case without a
central potential \cite{15} and generalized it to the relativistic
case \cite{16}.

Here we are interested in the more general cyclic solutions of the
alkaline atomic electron in the strong rotating magnetic field. As
pointed out in Ref. \cite{15}, these are available if the parameters
of the system are appropriately chosen. However, the nonadiabatic
geometric phases for such solutions were not calculated there. These
are now calculated in Sec. IV. The two solid angles subtended by the
traces of the orbit and spin angular momenta are also calculated
explicitly. It turns out that the nonadiabatic geometric phase in
this case also contains extra terms in addition to the linear
combination of the two solid angles.

In Sec. V we consider the alkaline atomic electron moving in an
arbitrarily varying strong magnetic field. We prove that the
nonadiabatic geometric phase for cyclic solutions with special
initial conditions is a linear combination of the two solid angles.
In other words, no extra term appears.

A brief summary is given in Sec. VI. A formula used in the text is
proved in the appendix.

\section{Neutral particles in a rotating magnetic field}   

Before the calculations begin, let us remark here some differences
between spin $1/2$ and higher ones. First, for any state of spin
$1/2$, say, an initial state $\Psi_0$, one can always find a unit
vector ${\bf e}_0$ such that ${\bf s}\cdot{\bf
e}_0\Psi_0=(1/2)\Psi_0$. In fact, ${\bf e}_0=2(\Psi_0,{\bf s}\Psi_0)$
is the unit vector to be found. From this fact and the result of Sec.
III, the previous conclusion for spin $1/2$ that the nonadiabatic
geometric phase is always proportional to the solid angle follows
immediately. For higher spin, on the other hand, the situation is
rather different. For a given state $\Psi_0$, in general one cannot
find a unit vector ${\bf e}_0$ such that ${\bf s}\cdot{\bf
e}_0\Psi_0=m_s\Psi_0$ ($m_s=s, s-1,\ldots, -s$). Let us give a simple
example for spin $3/2$. We denote the eigenstate of $s_z$ as
$\chi^0_{m_s}$, with eigenvalue $m_s$. Now consider the state $\Psi_0
= a \chi^0_{3/2} + b \chi^0_{-3/2}$, where $|a|^2 + |b|^2= 1$ for
normalization. The mean value of the spin in this state is $(\Psi_0,
{\bf s}\Psi_0)= (3 |a|^2 -3/2){\bf e}_z$ where ${\bf e}_z$ is the
unit vector in the $z$ direction. By varying $a$, the absolute value
of the above mean value may take any real number in the interval $[0,
3/2]$. Suppose that one could find a unit vector ${\bf e}_0$ such
that ${\bf s}\cdot{\bf e}_0\Psi_0=m_s\Psi_0$ ($m_s = \pm3/2, \pm
1/2$), then the mean value of the spin would be $(\Psi_0, {\bf
s}\Psi_0)= m_s{\bf e}_0$, and the absolute value is $|m_s|$, which is
obviously in contradiction with the above one. Second, even if the
mean value of ${\bf s}$ in $\Psi_0$ is specified, say, $(\Psi_0, {\bf
s}\Psi_0)=m_s{\bf e}_z$, one cannot assert that $s_z\Psi_0=m_s\Psi_0$
(the inverse is of course true) unless $m_s=\pm s$ (this is
automatically true for spin $1/2$). For example, for spin $3/2$, we
have infinitely many states $\chi=\chi^0_{1/2}$ and $\chi'
=e^{i\delta_1}\sqrt{2/3}\chi^0_{3/2}+e^{i\delta_2}(1/\sqrt
3)\chi^0_{-3/2}$ that lead to the mean value $\langle{\bf
s}\rangle={\bf e}_z/2$, where $\delta_1$ and $\delta_2$ are arbitrary
real numbers.

Consider a uniform magnetic field ${\bf B}(t)$ that has a constant
magnitude $B$ and rotates around some fixed axis at a constant angle
$\theta_B$ and with a constant frequency $\omega$. The rotating axis
is chosen as the $z$ axis of the coordinate system, so the magnetic
field is
\begin{equation}\label{1}
{\bf B}(t)=B{\bf n}(t), \quad {\bf n}(t)=(\sin\theta_B\cos\omega
t,\;\sin\theta_B\sin\omega t,\; \cos\theta_B)
\end{equation}
where $B$ and $\omega$ are taken to be positive without loss of
generality. Then consider a neutral particle with spin $s$
($s=1/2,1,3/2,\ldots$) and magnetic moment ${\bbox\mu}=\mu{\bf s}/s$,
where ${\bf s}$ is the spin operator in the unit of $\hbar$,
satisfying $[s_i,s_j] =i\epsilon_{ijk}s_k$. In the above magnetic
field, it has the time-dependent Hamiltonian
\begin{equation}\label{2}
H(t)=-{\bbox\mu}\cdot{\bf B}(t)=-\epsilon(\mu)\hbar\omega_B{\bf
s\cdot n}(t),
\end{equation}
where $\omega_B=|\mu|B/s\hbar$ is positive and $\epsilon(\mu)$ is the
sign function. The motion is governed by the Schr\"odinger equation
\begin{equation}\label{3}
i\hbar\partial_t\Psi=H(t)\Psi.
\end{equation}
To solve this equation, we make a unitary transformation \cite{2}
\begin{equation}\label{4}
\Psi(t)=W(t)\psi(t), \quad W(t)=\exp(-i\omega t s_z),
\end{equation}
then $\psi(t)$ satisfies a Schr\"odinger-like equation
\begin{equation}\label{5}
i\hbar\partial_t\psi=H_{\text{eff}}\psi,
\end{equation}
where the effective Hamiltonian reads
\begin{equation}\label{6}
H_{\text{eff}}=H(0)-\hbar\omega s_z =-\epsilon(\mu)\hbar\omega_B{\bf
s\cdot n}(0)-\hbar\omega s_z.
\end{equation}
This effective Hamiltonian is time independent, so that the
Schr\"odinger-like equation (\ref{5}) is readily integrable. For the
following convenience, we define the new quantities
\begin{mathletters}\label{7}
\begin{equation}\label{7a}
\omega_S=[\omega_B^2+\omega^2+2\epsilon(\mu)\omega_B\omega
\cos\theta_B]^{1/2},
\end{equation}
\begin{equation}\label{7b}
\sin\theta_S={\omega_B\sin\theta_B\over\omega_S},\quad
\cos\theta_S={\omega_B\cos\theta_B+\epsilon(\mu)\omega\over\omega_S},
\end{equation}
\begin{equation}\label{7c}
{\bf n}_S=(\sin\theta_S,\; 0,\; \cos\theta_S).
\end{equation}
\end{mathletters}
In terms of these new quantities, we have
\begin{equation}\label{8}
H_{\text{eff}}=-\epsilon(\mu)\hbar\omega_S{\bf s}\cdot{\bf n_S}.
\end{equation}
Therefore the Schr\"odinger-like equation (\ref{5}) is solved as
\begin{equation}\label{9}
\psi(t)=U_{\text{eff}}(t)\psi(0),\quad
U_{\text{eff}}(t)=\exp[i\epsilon(\mu)\omega_S t\;{\bf s}\cdot{\bf
n}_S].
\end{equation}
With the obvious relation $\Psi(0)=\psi(0)$, the Schr\"odinger
equation (\ref{3}) is solved as
\begin{mathletters}\label{10}
\begin{equation}\label{10a}
\Psi(t)=U(t)\Psi(0),
\end{equation}
where
\begin{equation}\label{10b}
U(t)=W(t)U_{\text{eff}}(t)=\exp(-i\omega t
s_z)\exp[i\epsilon(\mu)\omega_S t\; {\bf s}\cdot{\bf n}_S].
\end{equation}
\end{mathletters}
If one begins with an initial state $\Psi(t_0)$ at the time $t_0$
[but note that the time dependence of the magnetic field is still
given by Eq. (\ref{1})], then the solution reads
\begin{equation}\label{11}
\Psi(t)=W(t)U_{\text{eff}}(t-t_0)W^\dagger(t_0)\Psi(t_0).
\end{equation}
Since the evolution operator involves no chronological product, it is
convenient for practical calculations. In the following discussions
we will take the initial time to be $t_0=0$ for convenience.

First of all let us calculate the mean value of ${\bf s}$ in an
arbitrary state. We define
\begin{equation}\label{12}
{\bf v}(t)={\bbox(}\Psi(t), {\bf s}\Psi(t){\bbox )},
\end{equation}
and denote ${\bf v}_0={\bf v}(0)$. Using Eq. (\ref{10}) we have
$$
{\bf v}(t)={\bbox(}\Psi(0), U^\dagger_{\text{eff}}(t)W^\dagger(t)
{\bf s}W(t)U_{\text{eff}}(t)\Psi(0){\bbox )}.
$$
It is not difficult to show that
\begin{equation}\label{13}
W^\dagger(t){\bf s}W(t)=\exp(i\omega t s_z){\bf s}\exp(-i\omega t
s_z)=(s_x\cos\omega t-s_y\sin\omega t,\; s_x\sin\omega
t+s_y\cos\omega t,\; s_z).
\end{equation}
The following formula was proved in the appendix.
\begin{eqnarray}\label{14}
U^\dagger_{\text{eff}}(t){\bf s}U_{\text{eff}}(t)
&=&\exp[-i\epsilon(\mu)\omega_S t\;{\bf s}\cdot{\bf n}_S]{\bf s}
\exp[i\epsilon(\mu)\omega_S t\;{\bf s} \cdot{\bf n}_S] \nonumber\\
&=& [{\bf s}-({\bf s} \cdot{\bf n}_S){\bf n}_S]
\cos[\epsilon(\mu)\omega_S t]-({\bf n}_S \times{\bf s})
\sin[\epsilon(\mu)\omega_S t]+({\bf s} \cdot{\bf n}_S){\bf n}_S.
\end{eqnarray}
Using these two formulas we can calculate ${\bf v}(t)$ once ${\bf
v}_0$ is given. Let us define,
\begin{equation}\label{15}
{\bf g}(t)=[{\bf v}_0-({\bf v}_0 \cdot{\bf n}_S){\bf n}_S]
\cos[\epsilon(\mu)\omega_S t] -({\bf n}_S \times{\bf v}_0)
\sin[\epsilon(\mu)\omega_S t]+({\bf v}_0 \cdot{\bf n}_S){\bf n}_S,
\end{equation}
which is an ordinary vector, not an operator. Then the result reads
\begin{equation}\label{16}
{\bf v}(t)={\bbox(}g_x(t)\cos\omega t-g_y(t)\sin\omega t,\;
g_x(t)\sin\omega t+g_y(t)\cos\omega t,\; g_z(t){\bbox )}.
\end{equation}
This is a rather complicated result, but the physical picture is
clear. We observe that the three terms in ${\bf g}(t)$ are
perpendicular to one another, and
\begin{equation}\label{17}
|{\bf v}_0-({\bf v}_0 \cdot{\bf n}_S){\bf n}_S|=|{\bf n}_S \times
{\bf v}_0|=\sqrt{|{\bf v}_0|^2-({\bf v}_0 \cdot{\bf n}_S)^2} \equiv
v_{0\perp},
\end{equation}
so we define three unit vectors orthogonal to one another:
\begin{equation}\label{18}
{\bf e}^S_x=[{\bf v}_0-({\bf v}_0 \cdot{\bf n}_S){\bf n}_S]
/v_{0\perp},\quad {\bf e}^S_y={\bf n}_S \times{\bf v}_0 /v_{0\perp},
\quad {\bf e}^S_z={\bf n}_S,
\end{equation}
which constitute a right-handed frame. In this frame ${\bf g}(t)$
takes the form
\begin{equation}\label{19}
{\bf g}(t)=v_{0\perp}\cos[\epsilon(\mu)\omega_S t]{\bf e}^S_x
-v_{0\perp} \sin[\epsilon(\mu)\omega_S t]{\bf e}^S_y+({\bf v}_0
\cdot{\bf n}_S){\bf e}^S_z.
\end{equation}
This is nothing different from Eq. (\ref{15}). However, in this form
it help us recognize the physical picture of the motion, and
obviously yields $|{\bf v}(t)|=|{\bf g}(t)|=|{\bf v}_0|$ as expected.
Now to get the vector ${\bf v}(t)$, one just rotates ${\bf v}_0$
around ${\bf e}^S_z={\bf n}_S$ through an angle
$-\epsilon(\mu)\omega_S t$ (positive angle corresponds to
anti-clockwise rotation) to get ${\bf g}(t)$, and then rotates ${\bf
g}(t)$ around ${\bf e}_z$ through an angle $\omega t$. The resulted
motion involves nutation as well as rotation, and the motion is not
periodic in general. Note that a cyclic state leads to a periodic
${\bf v}(t)$, but the inverse is not necessarily true. Therefore to
obtain a cyclic solution, one should first find a periodic ${\bf
v}(t)$. Two cases with periodic ${\bf v}(t)$ are available. First, if
the parameters of the system are such that $\omega_S/\omega$ is a
rational number, then both $\omega t$ and $\omega_S t$ may
simultaneously become integral multiples of $2\pi$ at some latter
time $T$, and we have ${\bf v}(T)={\bf v}_0$, independent of the
initial condition. Second, if the initial condition is such that
\begin{equation}\label{20}
{\bf v}_0=m_s{\bf n}_S, \quad  m_s=s,s-1,\ldots,-s,
\end{equation}
we have ${\bf g}(t)=m_s{\bf n}_S$, and
\begin{equation}\label{21}
{\bf v}(t)=m_s {\bbox (} \sin\theta_S\cos\omega t,\;
\sin\theta_S\sin\omega t,\;\cos\theta_S{\bbox)}.
\end{equation}
In this case ${\bf v}(t)$ only makes rotation. It is obviously
periodic. In the following we will see that both cases indeed
correspond to cyclic solutions. It seems that no other cyclic
solution can be found.

Cyclic solutions of the second kind (with special initial condition)
have been previously discussed in detail \cite{2}. For comparison we
briefly review the result. First we give the eigenstates of ${\bf
s}\cdot{\bf n}_S$. It is not difficult to show that
\begin{equation}\label{22}
{\bf s}\cdot{\bf n}_S=\exp(-i\theta_S s_y)s_z\exp(i\theta_S s_y).
\end{equation}
We denote the eigenstates of $s_z$ with eigenvalues $m_s$
($m_s=s,s-1,\ldots,-s$) as $\chi^0_{m_s}$. Then the eigenstates of
${\bf s}\cdot{\bf n}_S$ with eigenvalues $m_s$ are obviously
\begin{equation}\label{23}
\chi_{m_s}=\exp(-i\theta_S s_y)\chi_{m_s}^0=\sum_{m'_s}D^{s}_{m'_s
m_s}(0,\theta_S,0)\chi_{m'_s}^0,
\end{equation}
where the $D$'s are Wigner functions.

We take the initial state to be
\begin{equation}\label{24}
\Psi_{m_s}(0)=\chi_{m_s}.
\end{equation}
This leads to Eqs. (\ref{20}) and (\ref{21}). The solution is indeed
cyclic, and the nonadiabatic geometric phase in a period
$\tau=2\pi/\omega$ was found to be \cite{2}
\begin{equation}\label{29}
\gamma_{m_s}=-m_s\Omega_S, \quad \text{mod $2\pi$},
\end{equation}
where $\Omega_S=2\pi(1-\cos\theta_S)$. We denote the solid angle
subtended by the trace of ${\bf v}(t)$ by $\Omega_{\bf v}$. Because
${\bf v}(t)$ is given by Eq. (\ref{21}) in the present case, we have
$$
\Omega_{\bf v}=\epsilon(m_s)\Omega_S,\quad \text{mod $4\pi$},
$$
and consequently
\begin{equation}\label{30}
\gamma_{m_s}=-|m_s|\Omega_{\bf v}, \quad \text{mod $2\pi$}.
\end{equation}
Therefore the geometric nature of the result is quite obvious. It
should be remarked that the spin angular momentum precesses
synchronously with the magnetic field, but at a different angle with
the rotating axis.

Now we consider cyclic solutions of more general forms. If the
parameters of the system are such that $\omega_S/\omega$ is a
rational number, then all solutions are cyclic, as shown below. We
denote $\omega_S/\omega=K_S/K$, where $K_S$ and $K$ are natural
numbers, prime to each other. Let $T=K\tau$, then $\omega T=2\pi K$
and $\omega_S T=2\pi K_S$. In this case ${\bf v}(t)$ is periodic with
period $T$. An arbitrary initial condition can be written as
\begin{equation}\label{31}
\Psi(0)=\sum_{m_s} c_{m_s}\chi_{m_s},
\end{equation}
where the coefficients $c_{m_s}$ are arbitrary except satisfying
$\sum_{m_s} |c_{m_s}|^2=1$ such that $\Psi(0)$ is normalized. Note
that $\Psi(0)$ can also be expanded in terms of the complete set
$\{\chi^0_{m_s}\}$, it is easy to find that
\begin{mathletters}\label{32}
\begin{equation}\label{32a}
\Psi(T)=\exp(i\delta)\Psi(0),
\end{equation}
where
\begin{equation}\label{32b}
\delta=s[\epsilon(\mu)2\pi K_S-2\pi K],\quad \text{mod $2\pi$}.
\end{equation}
\end{mathletters}
Therefore the state is indeed cyclic, and $\delta$ is the total phase
change which is independent of the initial condition. Using the
relation
\begin{equation}\label{26}
W^\dagger(t)H(t)W(t)=H(0)=H_{\text{eff}}+\hbar\omega s_z,
\end{equation}
and note that $H_{\text{eff}}$ commutes with $U_{\text{eff}}(t)$,
$s_z$ commutes with $W(t)$, we have
$$
\langle H(t)\rangle={\bbox(}\Psi(0), H_{\text{eff}}\Psi(0){\bbox )} +
\hbar\omega v_z(t) =-\epsilon(\mu)\hbar\omega_S{\bf v}_0\cdot{\bf
n}_S +\hbar\omega v_z(t).
$$
Thus the dynamic phase is
\begin{equation}\label{34}
\beta=-\hbar^{-1}\int_0^T \langle H(t)\rangle\;dt=\epsilon(\mu)2\pi
K_S{\bf v}_0\cdot{\bf n}_S-2\pi K\cos\theta_S {\bf v}_0\cdot{\bf
n}_S.
\end{equation}
This depends on the initial condition as expected. Finally we obtain
the nonadiabatic geometric phase
\begin{equation}\label{35}
\gamma=\delta-\beta=\epsilon(\mu)2\pi K_S(s-{\bf v}_0\cdot{\bf n}_S)
-2\pi K(s-\cos\theta_S {\bf v}_0\cdot{\bf n}_S),\quad \text{mod
$2\pi$}.
\end{equation}
In the special case when ${\bf v}_0=m_s {\bf n}_S$ this is consistent
with the previous result (note that $T=K\tau$).

The next task is to calculate geometrically the solid angle
$\Omega_{\bf v}$ subtended by the trace of ${\bf v}(t)$, and compare
it with $\gamma$. It is easy to show that
\begin{equation}\label{36}
\Omega_{\bf v}={1 \over |{\bf v}_0|}\int_0^T {v_x(t)\dot
v_y(t)-v_y(t)\dot v_x(t) \over |{\bf v}_0|+v_z(t)}\; dt.
\end{equation}
Because of the complicated results (\ref{15}) and (\ref{16}), it
would be difficult to calculate this straightforwardly. Let us try to
get around the difficulty. It is easy to show that
\begin{equation}\label{37}
v_x(t)\dot v_y(t)-v_y(t)\dot v_x(t)=g_x(t)\dot g_y(t)-g_y(t)\dot
g_x(t)+\omega[g_x^2(t)+g_y^2(t)].
\end{equation}
On account of the relations $|{\bf v}(t)|=|{\bf g}(t)|=|{\bf v}_0|$
and $g_z(t)=v_z(t)$, we have $g_x^2(t)+g_y^2(t)=|{\bf
v}_0|^2-v_z^2(t)$. Therefore
\begin{equation}\label{38}
\Omega_{\bf v}={1 \over |{\bf g}_0|}\int_0^T {g_x(t)\dot
g_y(t)-g_y(t)\dot g_x(t) \over |{\bf g}_0|+g_z(t)}\; dt +{1 \over
|{\bf v}_0|}\int_0^T \omega[|{\bf v}_0|-v_z(t)]\; dt,
\end{equation}
where ${\bf g}_0\equiv {\bf g}(0)={\bf v}_0$. The second integral can
be calculated easily, and the first is recognized as the solid angle
subtended by the trace of ${\bf g}(t)$, which is very easy to
calculate in the coordinate frame expanded by ${\bf e}^S_x$, ${\bf
e}^S_y$ and ${\bf e}^S_z$ [cf Eq. (\ref{19})]. The final result is
\begin{equation}\label{39}
\Omega_{\bf v}=-\epsilon(\mu)2\pi K_S\left(1-{{\bf v}_0\cdot{\bf
n}_S\over |{\bf v}_0|}\right)+2\pi K\left(1-\cos\theta_S{{\bf
v}_0\cdot{\bf n}_S\over |{\bf v}_0|}\right).
\end{equation}
Here the first term is due to the rotation around ${\bf e}^S_z={\bf
n}_S$, and the second is due to the further rotation around ${\bf
e}_z$. Compared with Eq. (\ref{35}), we find the relation
\begin{equation}\label{40}
\gamma=-|{\bf v}_0|\Omega_{\bf v}+(s-|{\bf v}_0|) [\epsilon(\mu)2\pi
K_S-2\pi K],\quad \text{mod $2\pi$}.
\end{equation}
Therefore $\gamma$ contains two terms. The first is the familiar one
that is proportional to $\Omega_{\bf v}$. The second is an extra
term. If $s=1/2$, it is easy to show that $|{\bf v}_0|=1/2$ for any
initial state, then the extra term vanishes automatically, and the
above relation reduces to $\gamma=-(1/2) \Omega_{\bf v}$, which is
known to be valid in an arbitrary magnetic field \cite{13,14}. For
higher spin, $s-|{\bf v}_0|$ is in general not an integer, and the
extra term cannot be dropped. For the special initial condition
(\ref{24}), the above relation reduces to the result (\ref{30}). We
will show in the next section that Eq. (\ref{30}) holds in an
arbitrarily varying magnetic field as long as the initial state is an
eigenstate of ${\bf s}\cdot{\bf e}_0$ with eigenvalue $m_s$, where
${\bf e}_0$ is some unit vector. For the rotating magnetic field at
hand, Eq. (\ref{30}) holds as long as $|{\bf v}_0|=m_s$. This is a
looser restriction on the initial condition. We do not know whether
this is true in a more general magnetic field.

To conclude this section we remark that the relation (\ref{40}) holds
when $|{\bf v}_0|=0$. This can be easily verified by comparing Eq.
(\ref{40}) with Eq. (\ref{35}) in this special case. Moreover, from
Eq. (\ref{34}) we see that the dynamic phase vanishes. Therefore one
may regard the total phase in this case as pure geometric, though
$\Omega_{\bf v}$ is not well defined.

\section{Neutral particles in an arbitrarily varying
magnetic field}

As seen in the last section, the relation $\gamma\propto\Omega_{\bf
v}$ does not always hold for spin higher than $1/2$. Thus it may be
of interest to ask when it would be valid in an arbitrarily varying
magnetic field. In this section we will show that a sufficient
condition is that the initial state is an eigenstate of ${\bf s}\cdot
{\bf e}_0$ where ${\bf e}_0$ is some unit vector.

As discussed at the beginning of Sec. II, given an arbitrary state
$\Psi(t)$ of spin $1/2$, one can always find a unit vector ${\bf
e}(t)$ such that ${\bf s}\cdot{\bf e}(t)\Psi(t)=(1/2)\Psi(t)$. This
holds at all times. For higher spin, however, no similar conclusion
is available. Nevertheless, we will show that if an eigenvalue
equation ${\bf s}\cdot{\bf e}_0\Psi(0)=m_s\Psi(0)$ holds for the
initial state $\Psi(0)$, a similar one with some appropriate unit
vector ${\bf e}(t)$ would hold at all later times. The latter
equation is of crucial importance since it enables us to explicitly
determine the state $\Psi(t)$ in terms of ${\bf e}(t)$ up to a phase
factor. If ${\bf e}(t)$ returns to ${\bf e}(0)$ at some later time
$T$, we obtain a cyclic solution.

We write down the Schr\"odinger equation in an arbitrarily varying
magnetic field ${\bf B}(t)=B(t){\bf n}(t)$:
\begin{equation}\label{41}
i\hbar\partial_t\Psi=H(t)\Psi=-\hbar\omega_B(t){\bf s}\cdot {\bf
n}(t)\Psi.
\end{equation}
There are two differences from the one in Sec. II. First, here
$\omega_B(t)=\mu B(t)/s\hbar$ is time dependent, and its sign may
changes with time [so we do not use $|\mu|$ in defining
$\omega_B(t)$]. Second, the unit vector ${\bf n}(t)$ is not given by
Eq. (\ref{1}), but varies arbitrarily. We would assume that the
magnetic field varies continuously.

We take the initial state $\Psi(0)$ of the system to be an eigenstate
of ${\bf s} \cdot{\bf e}_0$ with eigenvalue $m_s$ where ${\bf e}_0$
is some unit vector, that is
\begin{equation}\label{42}
{\bf s}\cdot{\bf e}_0\Psi(0)=m_s\Psi(0),\quad m_s=s,s-1,\ldots,-s.
\end{equation}
Let us define a vector ${\bf e}(t)$ by the following differential
equation and initial condition.
\begin{equation}\label{43}
\dot {\bf e}(t)=-\omega_B(t){\bf n}(t)\times {\bf e}(t),\quad {\bf
e}(0)={\bf e}_0.
\end{equation}
Obviously, $|{\bf e}(t)|$ is time independent, so ${\bf e}(t)$ is a
unit vector at any time. We are going to prove that
\begin{equation}\label{44}
{\bf s}\cdot{\bf e}(t)\Psi(t)=m_s\Psi(t)
\end{equation}
holds at all later times. This can be easily done by induction.

By definition, Eq. (\ref{44}) is valid at $t=0$. We assume that it is
valid at time $t$, what we need to do is to show that it is also true
at time $t+\Delta t$ where $\Delta t$ is an infinitesimal increment
of time. In fact, using Eqs. (\ref{41}) and (\ref{43}) we have
\begin{mathletters}\label{45}
\begin{equation}\label{45a}
\Psi(t+\Delta t)=\Psi(t)+i\omega_B(t){\bf s}\cdot{\bf n}(t)
\Psi(t)\Delta t,
\end{equation}
\begin{equation}\label{45b}
{\bf e}(t+\Delta t)={\bf e}(t)-\omega_B(t){\bf n}(t)\times{\bf e}(t)
\Delta t.
\end{equation}
\end{mathletters}
After some simple algebra, the conclusion is achieved.

Because ${\bf e}(t)$ is a unit vector, we can write
\begin{equation}\label{46}
{\bf e}(t)={\bbox (}\sin\theta_e(t)\cos\phi_e(t),\;
\sin\theta_e(t)\sin\phi_e(t),\; \cos\theta_e(t){\bbox )}.
\end{equation}
It is not difficult to show that
\begin{equation}\label{47}
{\bf s}\cdot{\bf e}(t)=\exp[-i\phi_e(t)s_z]\exp[-i\theta_e(t)s_y] s_z
\exp[i\theta_e(t)s_y]\exp[i\phi_e(t)s_z].
\end{equation}
Therefore the eigenstate of ${\bf s}\cdot{\bf e}(t)$ with eigenvalue
$m_s$ is
\begin{equation}\label{48}
\Psi(t)=\exp[i\alpha(t)]\exp[-i\phi_e(t)s_z]\exp[-i\theta_e(t)s_y]
\chi^0_{m_s},
\end{equation}
where $\alpha(t)$ is a phase that cannot be determined by the
eigenvalue equation. However, $\alpha(t)$ is not arbitrary. To
satisfy the Schr\"odinger equation, it should be determined by the
other variables $\theta_e(t)$ and $\phi_e(t)$. In fact, the above
equation yields
\begin{equation}\label{49}
{\bbox(}\Psi(t),\Psi(t+\Delta t){\bbox)}=1+i\dot\alpha(t)\Delta t
-im_s\cos[\theta_e(t)]\dot\phi_e(t)\Delta t.
\end{equation}
On the other hand, from Eq. (\ref{45a}) we have
\begin{equation}\label{50}
{\bbox(}\Psi(t),\Psi(t+\Delta t){\bbox)}=1+i\omega_B(t){\bf v}(t)
\cdot{\bf n}(t)\Delta t,
\end{equation}
where ${\bf v}(t)$ is defined by Eq. (\ref{12}). Comparing the two
results we obtain
\begin{equation}\label{51}
\dot\alpha(t)=m_s\cos[\theta_e(t)]\dot\phi_e(t)+\omega_B(t){\bf v}(t)
\cdot{\bf n}(t).
\end{equation}

The motion of ${\bf e}(t)$ is determined by the magnetic field. If
the magnetic field is such that ${\bf e}(t)$ returns to its initial
value at the time $T$, that is
\begin{equation}\label{52}
\theta_e(T)=\theta_e(0),\quad \phi_e(T)=\phi_e(0)+2\pi K,
\end{equation}
where $K$ is an integer, then we get a cyclic solution. In fact, it
is easy to show that
\begin{mathletters}\label{53}
\begin{equation}\label{53a}
\Psi(T)=\exp(i\delta)\Psi(0),
\end{equation}
where
\begin{equation}\label{53b}
\delta=\alpha(T)-\alpha(0)-2\pi m_s K, \quad \text{mod $2\pi$}
\end{equation}
\end{mathletters}
is the total phase change. Using Eqs. (\ref{51}) and (\ref{52}), it
can be recast in the form
\begin{equation}\label{54}
\delta=-m_s\int_0^T[1-\cos\theta_e(t)]\dot\phi_e(t)\;dt+\int_0^T
\omega_B(t){\bf v}(t)\cdot{\bf n}(t)\;dt.
\end{equation}
The second term is obviously the dynamic phase $\beta$. Therefore the
nonadiabatic geometric phase is
\begin{equation}\label{55}
\gamma=-m_s\Omega_{\bf e},\quad \text{mod $2\pi$}
\end{equation}
where
\begin{equation}\label{56}
\Omega_{\bf e}=\int_0^T[1-\cos\theta_e(t)]\dot\phi_e(t)\;dt
\end{equation}
is the solid angle subtended by the trace of ${\bf e}(t)$.

Finally notice that ${\bf v}(t)$ satisfies the same equation as ${\bf
e}(t)$, and ${\bf v}_0=m_s{\bf e}_0$ which can be easily verified, we
have ${\bf v}(t)=m_s{\bf e}(t)$. Consequently, Eq. (\ref{55}) can be
recast in the form
\begin{equation}\label{58}
\gamma=-|m_s|\Omega_{\bf v},\quad \text{mod $2\pi$}.
\end{equation}
This is the final result of this section. Though $\gamma$ and
$\Omega_{\bf v}$ cannot be explicitly calculated, the above relation
holds regardless of the form of the magnetic field. For $m_s=\pm s$,
this result was previously obtained in Ref. \cite{layton}, and for
general $m_s$ in Ref. \cite{gao}, both by different methods from
ours, but our method seems more straightforward and simpler.

It should be noted that ${\bf v}(t)=0$ when $m_s=0$. In this case
$\Omega_{\bf v}$ is not well defined. However, the final result
(\ref{58}) remains correct because it gives the same result
$\gamma=0$ as given by Eq. (\ref{55}). This remark also applies to
the result (\ref{30}) in Sec. II and similar ones in the following
sections. To conclude this section, we remark that for any cyclic
motion one can always appropriately choose the coordinate axes such
that $\theta_e(t)$ does not take on the values $0$ or $\pi$ during
the cycle under consideration. This avoids any uncontinuous jump or
ill definition of $\phi_e(t)$, and renders the above demonstration
sound enough.

\section{Alkaline atomic electron in a strong rotating
magnetic field} 

In this section we consider the valence electron of the alkaline atom
or that of the hydrogen atom moving in a strong rotating magnetic
field given by Eq. (\ref{1}). This is described by the Schr\"odinger
equation
\begin{mathletters}\label{59}
\begin{equation}\label{59a}
i\hbar\partial_t\Psi=H(t)\Psi,
\end{equation}
where
\begin{equation}\label{59b}
H(t)=H_0+\mu_{\rm B}B({\bf l}+2{\bf s})\cdot{\bf n}(t)
=H_0+\hbar\omega_B({\bf l}+2{\bf s})\cdot{\bf n}(t),
\end{equation}
in which
\begin{equation}\label{59c} H_0={{\bf p}^2\over 2M}+V(r)
\end{equation}
\end{mathletters}
is the Hamiltonian of the electron in the central potential of the
nucleus (and the other electrons in the inner shells for alkaline
atoms), $M$ and $\mu_{\rm B}$ are respectively the reduced mass and
the Bohr magneton of the electron, $\omega_B=\mu_{\rm B}B/\hbar>0$,
${\bf l}={\bf r}\times{\bf p}/\hbar$ is the orbit angular momentum
operator in unit of $\hbar$, and ${\bf s}$ the spin as before (here
$s=1/2$). The applicability of this equation was discussed in Ref.
\cite{15}.

The above Schr\"odinger equation can be solved in a way similar to
that in Sec. II. This was discussed in detail in Ref. \cite{15}. The
solution is
\begin{equation}\label{60}
\Psi(t)=U(t)\Psi(0),\quad U(t)=W(t)U_{\text{eff}}(t),
\end{equation}
where
\begin{equation}\label{61}
W(t)=\exp(-i\omega t j_z),\quad U_{\rm eff}(t)=\exp(-iH_{\text{
eff}}t/\hbar),
\end{equation}
where $j_z$ is the $z$-component of the total angular momentum (in
unit of $\hbar$) ${\bf j=l+s}$, and
\begin{equation}\label{62}
H_{\text{eff}}=H_0+\hbar\omega_L{\bf l}\cdot{\bf n}_L +\hbar\omega_S
{\bf s}\cdot{\bf n}_S
\end{equation}
is the effective Hamiltonian. The parameters in the effective
Hamiltonian are defined as
\begin{mathletters}\label{63}
\begin{equation}\label{63a}
\omega_L=(\omega_B^2+\omega^2-2\omega_B\omega\cos\theta_B)^{1/2},
\end{equation}
\begin{equation}\label{63b}
\omega_S=(4\omega_B^2+\omega^2-4\omega_B\omega\cos\theta_B)^{1/2};
\end{equation}
\end{mathletters}
\begin{equation}\label{64}
{\bf n}_L=(\sin\theta_L,\; 0,\; \cos\theta_L),\quad {\bf n}_S
=(\sin\theta_S,\; 0,\; \cos\theta_S),
\end{equation}
where
\begin{mathletters}\label{65}
\begin{equation}\label{65a}
\sin\theta_L={\omega_B\sin\theta_B\over\omega_L},\quad
\cos\theta_L={\omega_B\cos\theta_B-\omega\over\omega_L},
\end{equation}
\begin{equation}\label{65b}
\sin\theta_S={2\omega_B\sin\theta_B\over\omega_S},\quad
\cos\theta_S={2\omega_B\cos\theta_B-\omega\over\omega_S}.
\end{equation}
\end{mathletters}

Using the above solution we can calculate the mean values of ${\bf
l}$ and ${\bf s}$ in an arbitrary state. We define
\begin{equation}\label{66}
{\bf u}(t)={\bbox(}\Psi(t), {\bf l}\Psi(t){\bbox )},\quad {\bf
v}(t)={\bbox(}\Psi(t), {\bf s}\Psi(t){\bbox )},
\end{equation}
and denote ${\bf u}_0={\bf u}(0)$ and ${\bf v}_0={\bf v}(0)$. On
account of the fact that ${\bf l}$, ${\bf s}$ and $H_0$ commute with
one another, the results can be obtained by calculations similar to
those performed in Sec. II. We define
\begin{mathletters}\label{67}
\begin{equation}\label{67a}
{\bf f}(t)=[{\bf u}_0-({\bf u}_0 \cdot{\bf n}_L){\bf n}_L]
\cos\omega_L t +({\bf n}_L \times{\bf u}_0) \sin\omega_L t+({\bf u}_0
\cdot{\bf n}_L){\bf n}_L,
\end{equation}
\begin{equation}\label{67b}
{\bf g}(t)=[{\bf v}_0-({\bf v}_0 \cdot{\bf n}_S){\bf n}_S]
\cos\omega_S t +({\bf n}_S \times{\bf v}_0) \sin\omega_S t+({\bf v}_0
\cdot{\bf n}_S){\bf n}_S,
\end{equation}
\end{mathletters}
then the results read
\begin{mathletters}\label{68}
\begin{equation}\label{68a}
{\bf u}(t)={\bbox(}f_x(t)\cos\omega t-f_y(t)\sin\omega t,\;
f_x(t)\sin\omega t+f_y(t)\cos\omega t,\; f_z(t){\bbox )},
\end{equation}
\begin{equation}\label{68b}
{\bf v}(t)={\bbox(}g_x(t)\cos\omega t-g_y(t)\sin\omega t,\;
g_x(t)\sin\omega t+g_y(t)\cos\omega t,\; g_z(t){\bbox )}.
\end{equation}
\end{mathletters}
Note that the second term in ${\bf g}(t)$ has a different sign from
the one in Sec. II. The physical picture is rather similar to the
previous one.

We denote the common eigenstates of $\{H_0,\;{\bf l}^2,\;l_z\}$ as
$\zeta_{nlm}^0$, with eigenvalues $\{\epsilon_{nl},\; l(l+1),\; m\}$,
where $\epsilon_{nl}$ are the energy spectrum of the electron in the
absence of the magnetic field. Then the common eigenstates of the
operators $\{H_0,\;{\bf l}^2,\;{\bf l}\cdot{\bf n}_L,\;{\bf s}\cdot
{\bf n}_S\}$ with eigenvalues $\{\epsilon_{nl},\; l(l+1),\; m,\;
m_s\}$ (now $m_s=\pm 1/2$) are
\begin{equation}\label{69}
\varphi_{nlmm_s}=\zeta_{nlm}\chi_{m_s},
\end{equation}
where $\chi_{m_s}$ is given by Eq. (\ref{23}), and
\begin{equation}\label{70}
\zeta_{nlm}=\exp(-i\theta_L l_y)\zeta_{nlm}^0
=\sum_{m'}D^l_{m'm}(0,\theta_L,0)\zeta_{nlm'}^0.
\end{equation}
The $\varphi_{nlmm_s}$ are also eigenstates of $H_{\text{eff}}$ with
the eigenvalues
\begin{equation}\label{71}
E_{nlmm_s}=\epsilon_{nl}+m\hbar\omega_L+m_s\hbar\omega_S.
\end{equation}
However, these eigenvalues are not observable since $H_{\text{eff}}$
is not a physical quantity. The above states are complete. At a given
time, any state of the system can be expressed as a linear
combination of them.

As shown in Ref. \cite{15}, a solution with the initial condition
\begin{equation}\label{72}
\Psi_i(0)=\varphi_i=\varphi_{nlmm_s}
\end{equation}
is a cyclic one. Here for convenience we use one subscript $i$ to
represent all the quantum numbers $nlmm_s$. The nonadiabatic
geometric phase was shown to be
\begin{equation}\label{73}
\gamma_i=-m\Omega_L-m_s\Omega_S, \quad \text{mod $2\pi$}.
\end{equation}
Here $\Omega_L=2\pi(1-\cos\theta_L)$ and
$\Omega_S=2\pi(1-\cos\theta_S)$. We denote the solid angles subtended
by the traces of the orbit and spin angular momenta by $\Omega_{\bf
u}$ and $\Omega_{\bf v}$, respectively. For the present case ${\bf
u}(t)$ and ${\bf v}(t)$ have forms similar to Eq. (\ref{21}), so we
have
\begin{equation}\label{75}
\gamma_{i}=-|m|\Omega_{\bf u}-|m_s|\Omega_{\bf v}, \quad \text{mod
$2\pi$},
\end{equation}
and the geometric nature of the result is quite obvious. For the
alkaline atomic electron at hand, $|m_s|$ can be replaced by $1/2$
since $m_s=\pm 1/2$. However, the above result is valid for a charged
particle of general spin $s$ moving in the central potential plus the
strong rotating magnetic field. In Sec. V we will show that this
result holds in an arbitrarily varying strong magnetic field as long
as the initial state is appropriately chosen. However, we will show
in the following that for more general cyclic solutions the
nonadiabatic geometric phase contains extra terms in addition to the
linear combination of the two solid angles.

In the above we see that the initial condition
$\Psi_{i}(0)=\varphi_i$ leads to a cyclic solution in the general
case. If the parameters of the system satisfy some appropriate
conditions, more cyclic solutions are available. In fact, if
$\omega_B$, $\omega$ and $\theta_B$ are such that both
$\omega_L/\omega$ and $\omega_S/\omega$ are rational numbers, we will
see that any solution with the initial condition
\begin{equation}\label{76}
\Psi(0)=\sum_{mm_s}c_{mm_s}\varphi_{nlmm_s}
\end{equation}
is a cyclic one, where the coefficients $c_{mm_s}$ are arbitrary
except satisfying $\sum_{mm_s}|c_{mm_s}|^2=1$ such that the initial
state is normalized. Suppose that $\omega_L/\omega=K_2/K_1$,
$\omega_S/\omega=K_4/K_3$, where all the $K$'s are natural numbers,
with $K_2$ and $K_1$ prime to each other, and the same for $K_4$ and
$K_3$. We denote the least common multiple of $K_1$ and $K_3$ as $K$,
and write
\begin{equation}\label{77}
\omega_L/\omega=K_L/K, \quad \omega_S/\omega=K_S/K,
\end{equation}
where $K_L=KK_2/K_1$ and $K_S=KK_4/K_3$ are both natural numbers. In
this case, $\omega T$, $\omega_L T$ and $\omega_S T$ are all integral
multiples of $2\pi$, where $T=K\tau$ is now the period of ${\bf
u}(t)$ and ${\bf v}(t)$. It is not difficult to show that
\begin{mathletters}\label{78}
\begin{equation}\label{78a}
\Psi(T)=\exp(i\delta)\Psi(0),
\end{equation}
where
\begin{equation}\label{78b}
\delta=-\epsilon_{nl}T/\hbar-l(2\pi K+2\pi K_L) -s(2\pi K+2\pi
K_S),\quad \text{mod $2\pi$}.
\end{equation}
\end{mathletters}
Thus the solution is actually cyclic and $\delta$ is the total phase
change which is independent of the initial condition. Currently we
would like to keep the general value $s$ such that the result may be
applied to a charged particle with more general spin. The dynamic
phase can be calculated in a way similar to that in Sec. II, the
result is
\begin{equation}\label{79}
\beta=-\epsilon_{nl}T/\hbar-2\pi K_L{\bf u}_0\cdot{\bf n}_L-2\pi
K_S{\bf v}_0\cdot{\bf n}_S-2\pi K(\cos\theta_L {\bf u}_0\cdot{\bf
n}_L+\cos\theta_S {\bf v}_0\cdot{\bf n}_S).
\end{equation}
This depends on the initial condition as expected. The nonadiabatic
geometric phase is
\begin{eqnarray}\label{80}
\gamma=\delta-\beta=&&-2\pi K(l-\cos\theta_L{\bf u}_0\cdot{\bf n}_L)
-2\pi K(s-\cos\theta_S{\bf v}_0\cdot{\bf n}_S)\nonumber\\
&&-(l-{\bf u}_0\cdot{\bf n}_L)2\pi K_L-(s-{\bf v}_0\cdot{\bf
n}_S)2\pi K_S,\quad \text{mod $2\pi$}.
\end{eqnarray}
On the other hand, the solid angles subtended by the traces of the
orbit and spin angular momenta are
\begin{mathletters}\label{81}
\begin{equation}\label{81a}
\Omega_{\bf u}=2\pi K_L\left(1-{{\bf u}_0\cdot{\bf n}_L\over |{\bf
u}_0|}\right)+2\pi K\left(1-\cos\theta_L{{\bf u}_0\cdot{\bf n}_L\over
|{\bf u}_0|}\right),
\end{equation}
\begin{equation}\label{81b}
\Omega_{\bf v}=2\pi K_S\left(1-{{\bf v}_0\cdot{\bf n}_S\over |{\bf
v}_0|}\right)+2\pi K\left(1-\cos\theta_S{{\bf v}_0\cdot{\bf n}_S\over
|{\bf v}_0|}\right).
\end{equation}
\end{mathletters}
The calculations that lead to these results are similar to those in
Sec. II. From the above results and Eq. (\ref{80}) it is easy to find
that
\begin{equation}\label{82}
\gamma=-|{\bf u}_0|\Omega_{\bf u}-|{\bf v}_0|\Omega_{\bf v}+ (|{\bf
u}_0|-l)(2\pi K+2\pi K_L)+ (|{\bf v}_0|-s)(2\pi K+2\pi K_S),\quad
\text{mod $2\pi$}.
\end{equation}
Therefore, in addition to a linear combination of $\Omega_{\bf u}$
and $\Omega_{\bf v}$, we get two extra terms in $\gamma$. This result
holds for a charged particle with spin $s$ moving in the central
potential plus the magnetic field. The first extra term can be
dropped only when $|{\bf u}_0|$ is an integer and the second when
$|{\bf v}_0|-s$ is an integer. For the alkaline atomic electron,
$|{\bf v}_0|=s=1/2$, so the second extra term vanishes. Because $l$
is an integer, we have for the alkaline atomic electron the final
result
\begin{equation}\label{83}
\gamma=-|{\bf u}_0|\Omega_{\bf u}-{\textstyle\frac 12}\Omega_{\bf v}+
|{\bf u}_0|(2\pi K+2\pi K_L),\quad \text{mod $2\pi$}.
\end{equation}
For the special initial condition (\ref{72}), the above results
reduce to Eq. (\ref{75}). In the next section we will show that the
result (\ref{75}) holds in an arbitrarily varying magnetic field as
long as the initial state is a common eigenstate of $\{ H_0,\; {\bf
l}^2,\; {\bf l}\cdot{\bf d}_0,\; {\bf s}\cdot{\bf e}_0\}$ with
eigenvalues $\{\epsilon_{nl},\; l(l+1),\; m,\; m_s\}$, where ${\bf
d}_0$ and ${\bf e}_0$ are some unit vectors. For the rotating
magnetic field at hand, it holds as long as $|{\bf u}_0|=|m|$ and
$|{\bf v}_0|=|m_s|$ (the second is automatically satisfied for
$s=1/2$), which is a looser restriction on the initial state. We do
not know whether this is true for a more general magnetic field. To
conclude this section we point out that the condition (\ref{77}) can
be realized with $\omega_L/\omega=1$ and $\omega_S/\omega=2$, if one
chooses $\omega_B/\omega=\sqrt{3/2}$ and $\cos\theta_B=\sqrt 3/2\sqrt
2$ \cite{15}.

\section{Alkaline atomic electron in an arbitrarily varying
strong magnetic field} 

Now we consider the alkaline atomic electron moving in an arbitrarily
varying strong magnetic field. We write down the Schr\"odinger
equation:
\begin{mathletters}\label{84}
\begin{equation}\label{84a}
i\hbar\partial_t\Psi=H(t)\Psi,
\end{equation}
where
\begin{equation}\label{84b}
H(t)=H_0+\hbar\omega_B(t)({\bf l}+2{\bf s})\cdot{\bf n}(t).
\end{equation}
\end{mathletters}
Compared with that in Sec. IV, there are two differences. First,
$\omega_B(t)=\mu_{\rm B}B(t)/\hbar$ is time dependent and may be
either positive or negetive. Second, ${\bf n}(t)$ is an arbitrarily
varying unit vector. It is difficult to obtain any specific solution
of the above equation. However, for solutions of special initial
conditions, we can establish a relation between $\gamma$ and the
solid angles $\Omega_{\bf u}$ and $\Omega_{\bf v}$.

Obviously, the operators in the set $\{ H_0,\; {\bf l}^2,\; {\bf
l}\cdot{\bf d}_0,\; {\bf s}\cdot{\bf e}_0\}$ commute with one
another, where ${\bf d}_0$ and ${\bf e}_0$ are some unit vectors,
thus they can have a complete set of common eigenstates. We take the
initial state $\Psi(0)$ of the system to be such a common eigenstate,
that is
\begin{mathletters}\label{85}
\begin{equation}\label{85a}
H_0\Psi(0)=\epsilon_{nl}\Psi(0),\quad {\bf l}^2\Psi(0)=l(l+1)\Psi(0),
\end{equation}
\begin{equation}\label{85b}
{\bf l}\cdot{\bf d}_0\Psi(0)=m\Psi(0),\quad {\bf s}\cdot {\bf
e}_0\Psi(0)=m_s\Psi(0).
\end{equation}
\end{mathletters}
For the alkaline atomic electron, the last condition need not be
assumed, since one can always find a unit vector ${\bf e}_0$ such
that it holds with $m_s=1/2$ or $-1/2$. However, we prefer to assume
it so that the result may be valid for charged particles of more
general spin. We define two vectors ${\bf d}(t)$ and ${\bf e}(t)$ by
the following differential equations and initial conditions
\begin{mathletters}\label{86}
\begin{equation}\label{86a}
\dot {\bf d}(t)=\omega_B(t){\bf n}(t)\times {\bf d}(t),\quad {\bf
d}(0)={\bf d}_0,
\end{equation}
\begin{equation}\label{86b}
\dot {\bf e}(t)=2\omega_B(t){\bf n}(t)\times {\bf e}(t),\quad {\bf
e}(0)={\bf e}_0.
\end{equation}
\end{mathletters}
Obviously, both $|{\bf d}(t)|$ and $|{\bf e}(t)|$ are time
independent, so ${\bf d}(t)$ and ${\bf e}(t)$ are unit vectors at any
time. One can prove that the following eigenvalue equations hold at
all times.
\begin{mathletters}\label{87}
\begin{equation}\label{87a}
H_0\Psi(t)=\epsilon_{nl}\Psi(t),\quad {\bf l}^2\Psi(t)=l(l+1)\Psi(t),
\end{equation}
\begin{equation}\label{87b}
{\bf l}\cdot{\bf d}(t)\Psi(t)=m\Psi(t),\quad {\bf s}\cdot {\bf
e}(t)\Psi(t)=m_s\Psi(t).
\end{equation}
\end{mathletters}

Because ${\bf d}(t)$ and ${\bf e}(t)$ are unit vectors, we can write
\begin{mathletters}\label{88}
\begin{equation}\label{88a}
{\bf d}(t)={\bbox (}\sin\theta_d(t)\cos\phi_d(t),\;
\sin\theta_d(t)\sin\phi_d(t),\; \cos\theta_d(t){\bbox )},
\end{equation}
\begin{equation}\label{88b}
{\bf e}(t)={\bbox (}\sin\theta_e(t)\cos\phi_e(t),\;
\sin\theta_e(t)\sin\phi_e(t),\; \cos\theta_e(t){\bbox )}.
\end{equation}
\end{mathletters}
As in Sec. III, $\Psi(t)$ can be written as
\begin{equation}\label{89}
\Psi(t)=\exp[i\alpha(t)]\exp[-i\phi_d(t)l_z]\exp[-i\theta_d(t)l_y]
\zeta^0_{nlm} \exp[-i\phi_e(t)s_z]\exp[-i\theta_e(t)s_y]
\chi^0_{m_s},
\end{equation}
where $\alpha(t)$ is determined by the equation
\begin{equation}\label{90}
\dot\alpha(t)=m\cos[\theta_d(t)]\dot\phi_d(t)+
m_s\cos[\theta_e(t)]\dot\phi_e(t)-\hbar^{-1}\langle H(t)\rangle,
\end{equation}
where the expectation value $\langle H(t)\rangle$ is calculated in
the state $\Psi(t)$.

The motions of ${\bf d}(t)$ and ${\bf e}(t)$ are determined by the
magnetic field. If the latter is such that both ${\bf d}(t)$ and
${\bf e}(t)$ return to their initial values at time $T$, then we get
a cyclic solution. The nonadiabatic geometric phase can be shown to
be
\begin{equation}\label{94}
\gamma=-m\Omega_{\bf d}-m_s\Omega_{\bf e},\quad \text{mod $2\pi$}
\end{equation}
where
\begin{equation}\label{95}
\Omega_{\bf d}=\int_0^T[1-\cos\theta_d(t)]\dot\phi_d(t)\;dt, \quad
\Omega_{\bf e}=\int_0^T[1-\cos\theta_e(t)]\dot\phi_e(t)\;dt
\end{equation}
are the solid angles subtended by the traces of ${\bf d}(t)$ and
${\bf e}(t)$.

By similar reasoning to that in Sec. III, we have ${\bf u}(t)=m{\bf
d}(t)$ and ${\bf v}(t)=m_s{\bf e}(t)$, and Eq. (\ref{94}) can be
written as
\begin{equation}\label{97}
\gamma=-|m|\Omega_{\bf u}-|m_s|\Omega_{\bf v},\quad \text{mod
$2\pi$}.
\end{equation}
This is the final result of this section. It holds for a charged
particle with spin $s$ moving in the central potential plus the
arbitrarily varying strong magnetic field. For the alkaline atomic
electron, $|m_s|$ may be replaced by $1/2$.

\section{Summary}  

In this paper we have studied the nonadiabatic geometric phases of
neutral or charged particles moving in a time-dependent magnetic
field. In Sec. II we consider a neutral particle with general spin
and with magnetic moment moving in a rotating magnetic field. The
nonadiabatic geometric phase for special cyclic solutions is
proportional to the solid angle subtended by the trace of the spin
[cf Eq. (\ref{30})]. This is well known. However, for more general
cyclic solutions, we find that the nonadiabatic geometric phase
contains an extra term. The main result of this section is Eq.
(\ref{40}). The extra term vanishes automatically for spin $1/2$,
consistent with the known conclusion for spin $1/2$ that Eq.
(\ref{30}) is valid in an arbitrary magnetic field. For higher spin,
however, the extra term depends on the initial condition. In Sec. III
we show that the result (\ref{30}) is valid for special cyclic
solutions of higher spin in an arbitrarily varying magnetic field. In
Sec. IV we consider a charged particle moving in a central potential
plus a strong rotating magnetic field. This may describe the valence
electron in an alkaline atom or that in a hydrogen atom. For special
cyclic solutions, the nonadiabatic geometric phase is a linear
combination of the two solid angles subtended by the traces of the
orbit and spin angular momenta [cf Eq. (\ref{75})]. This is also a
previously known result. For the more general cyclic solutions,
however, extra terms are also involved in the geometric phase. The
main results of this section are Eqs. (\ref{82}) and (\ref{83}). In
Sec. V we prove that the result (\ref{75}) is valid for special
cyclic solutions of charged particles moving in a central potential
plus an arbitrarily varying strong magnetic field.

\section*{Acknowledgments}

This work was supported by the National Natural Science Foundation of
the People's Republic of China.

\newpage
\section*{Appendix}
\baselineskip 15pt

Here we prove Eq. (\ref{14}) in a very simple way.

We define
\begin{equation}\eqnum{A1}\label{A1}
{\bf F}(\phi)=\exp(i\phi\;{\bf s}\cdot{\bf n}_S){\bf s}
\exp(-i\phi\;{\bf s} \cdot{\bf n}_S).
\end{equation}
Differentiation of this equation with respect to $\phi$ yields
\begin{equation}\eqnum{A2}\label{A2}
{\bf F}'(\phi)={\bf n}_S\times {\bf F}(\phi),
\end{equation}
and
\begin{equation}\eqnum{A3}\label{A3}
{\bf F}''(\phi)={\bf n}_S\times [{\bf n}_S\times{\bf F}(\phi)] =[{\bf
F}(\phi)\cdot{\bf n}_S]{\bf n}_S-{\bf F}(\phi).
\end{equation}
From Eq. (\ref{A2}) we have $[{\bf F}(\phi)\cdot{\bf n}_S]'={\bf
F}'(\phi)\cdot{\bf n}_S=0$, so that ${\bf F}(\phi)\cdot{\bf n}_S=
{\bf F}(0)\cdot{\bf n}_S={\bf s}\cdot{\bf n}_S$. Then Eq. (\ref{A3})
becomes
\begin{equation}\eqnum{A4}\label{A4}
{\bf F}''(\phi)+{\bf F}(\phi)=({\bf s}\cdot {\bf n}_S){\bf n}_S.
\end{equation}
The solution of this equation is obviously
\begin{equation}\eqnum{A5}\label{A5}
{\bf F}(\phi)={\bf a}\cos\phi+{\bf b}\sin\phi+({\bf s}\cdot {\bf
n}_S){\bf n}_S,
\end{equation}
where ${\bf a}$ and ${\bf b}$ are constant vectors. From Eqs.
(\ref{A1}) and (\ref{A2}) we have
\begin{equation}\eqnum{A6}\label{A6}
{\bf F}(0)={\bf s}, \quad {\bf F}'(0)={\bf n}_S\times{\bf s}.
\end{equation}
This determines ${\bf a}$ and ${\bf b}$, so we arrive at
\begin{equation}\eqnum{A7}\label{A7}
{\bf F}(\phi)=[{\bf s}-({\bf s}\cdot {\bf n}_S){\bf n}_S]
\cos\phi+({\bf n}_S\times{\bf s})\sin\phi+({\bf s}\cdot {\bf
n}_S){\bf n}_S.
\end{equation}
Eq. (\ref{14}) can be obtained by substituting $\phi=-\epsilon(\mu)
\omega_S t$ into the above result.

\end{document}